\documentclass{aa}
\usepackage{txfonts,rotating}
\usepackage{graphicx}

\usepackage{color}

\begin{document}

\def\msun{\hbox{M$_\odot$}}
\def\noao{\hbox{The NOAO Deep Wide Field Survey MOSAIC Data Reductions, 
http://www.noao.edu/noao/noaodeep/ReductionOpt/frames.html}}
\def\prepare{\hbox{in preparation}}
\def\inpress{\hbox{ }}

\title{Pal\,13: its moderately extended low density halo and its accretion history}

\author{Andr\'es E. Piatti\inst{1,2}\thanks{\email{andres.piatti@unc.edu.ar}} and Jos\'e G. 
Fern\'andez-Trincado\inst{3}
}

\institute{Consejo Nacional de Investigaciones Cient\'{\i}ficas y T\'ecnicas, Godoy Cruz 2290, 
C1425FQB,  Buenos Aires, Argentina;
\and
Observatorio Astron\'omico, Universidad Nacional de C\'ordoba, Laprida 854, 5000, 
C\'ordoba, Argentina;
\and Instituto de Astronom\'{i}a y Ciencias Planetarias, Universidad de Atacama, Copayapu 485, 
Copiap\'o, Chile\\
}

\date{Received / Accepted}

\abstract{
We present results on the basis of Dark Energy Camera Legacy Survey 
(DECaLS) DR8 astrometric and photometric data sets of the Milky Way
globular cluster Pal\,13. Because of its relative small size and mass,
there has not been yet a general consensus about the existence of extra-tidal
structures around it. While some previous results claim for the absence of such
features, others have shown that the cluster is under the effects of tidal stripping.
From DECaLS $g,r$ magnitudes of stars placed along the cluster Main Sequence 
in the color-magnitude diagram --previously corrected by interstellar reddening--,
we built the cluster stellar density map. The resulting density map shows
nearly smooth  contours around Pal\,13 out to $\approx$ 1.6 times the most
recent estimate of its Jacobi radius, derived by taking into account its 
variation along its orbital motion. This outcome favors the presence of stars 
escaping the cluster, a phenomenon
frequently seen in globular clusters that have crossed the Milky Way disc
a comparable large number of times. Particularly, the orbital high
eccentricity and large inclination angle of this accreted globular cluster
could have been responsible for the relatively large amount
of cluster mass lost. 
}


\titlerunning{The Milky Way globular cluster Pal\,13}
\authorrunning{A.E. Piatti and J.G. Fern\'andez-Trincado}

\maketitle

\markboth{A.E. Piatti and J.G. Fern\'andez-Trincado: The Milky Way globular cluster Pal\,13}{}

\section{Introduction}
Extended stellar structures have been
observed around a non-negligible number of Galactic globular clusters \citep{carballobelloetal2012}. For instance, \citet{correntietal2011} discovered
an extended stellar halo surrounding the distant NGC\,5694, while 
 \citet{olszewskietal2009} found an unprecedented extra-tidal, azimuthally smooth, 
halo-like diffuse spatial extension of NGC\,1851. A similar structure was also
found around 47\,Tuc \citet{p17c}. Long tidal tails have  been detected in the field of 
NGC\,288 \citep{shippetal2018}, NGC\,5466 \citep{belokurovetal2006}, NGC\,7492 
\citep{naverreteetal2017}, Pal\,1 \citep{noetal2010}, Pal\,5 \citep{odenetal2003}, Pal\,14 
\citep{sollimaetal2011}, Pal\,15 \citep{myeongetal2017}, $\omega$ Cen \citep{ibataetal2019} 
and M\,5 \citep{g2019}; and other globular clusters
were found to be embedded in a diffuse stellar envelope extending to a radial distance 
of at least five time the nominal tidal radius, like M\,2 \citep{kuzmaetal2016}.
From a theoretical point of view, some N-body simulations  show
that potential escapers  \citep{kupperetal2010} or potential observational biases
\citep{bg2018} could contribute to the detection of extended  envelopes around 
globular clusters, among others.

There has been some discussion in the literature about the existence of 
extra-tidal features around Pal\,13. \citet{kupperetal2011} performed $N$-body 
simulations from which they found that the cluster is most likely near to its 
apogalacticon and therefore appear supervirial and blown-up, so that
extra-tidal stars got pushed back into the vicinity of the cluster. 
\citet{bradfordetal2011} obtained an outer surface brightness slope 
shallower than that for typical globular clusters, so that at large distance,
tidal debris are likely affecting the cluster stellar density profile. According to 
the authors, they could be an evidence for tidal stripping. Another couple of
still not reconciled results are those derived by \citet{kunduetal2019} and
\citet{yepezetal2019}, respectively, using {\it Gaia} proper motions. While the 
former concluded on the absence of extra-tidal  RR Lyrae stars ripped apart 
from the cluster due to tidal disruption, the latter found that {\it Gaia} proper 
motions of cluster members show a significant scatter, consistent with an
scenario of a cluster being tidally stripped. \citet{siegeletal2001}  also 
suggested a significant degree of tidal destruction on the basis of various 
observational evidence.

It is worth mentioning that the cluster tidal radius has been attained in several 
studies, resulting in remarkably different values. For instance, \citet{coteetal2002} 
determined a tidal radius of $r_t$ = 26$\arcmin$$\pm$6$\arcmin$ from surface 
density and surface brightness profiles. The authors mention that details of 
background subtraction and model-fitting lead to describe Pal\,13 as either 
containing a substantial population of  extra-tidal stars, or being considerably
more spatially extended than previously thought. Later, \citet{bradfordetal2011}
derived a smaller tidal radius of $r_t$ = 13.9$\arcmin$$\pm$1.5$\arcmin$ from a
maximum likelihood method applied to all stars in a color-magnitude
diagram (CMD) selection window, while \citet{sollimaetal2018} using the
same data set and a different analysis method adopted $r_{lim}$ = 
11.2$\arcmin$. More recently, \citet{baumgardtetal2019} estimated
$r_t$ = 4.94$\arcmin$ by comparing the cluster density profile to a large suite 
of direct $N$-body star cluster simulations. Note that the value tabulated
in \citet[][2010 Edition]{harris1996}'s catalogue is $r_t$ = 2.19$\arcmin$.

\citet{massarietal2019} have associated the origin of the globular cluster 
Pal\,13 to the {\it Sequoia} dwarf galaxy, which took part of an
early substantial accretion event that contributed to the formation of the
Milky Way stellar halo \citep{myeongetal2019}, which is also supported
by the identification of extended tidal debris with globular cluster abundance like
patterns in the inner halo \citep[e.g.][]{fetal2019a}. Other six globular
clusters have also been associated to the same progenitor, namely: 
FSR\,1758, IC\,4499, NGC\,3201, 5466, 6101 and 7006. As far as we are aware,
4 of them with previous studies of their external structures have been found to
have extra-tidal features : FSR\,1758:
\citet{barbaetal2019}; NGC\,3202: \citet{kunderetal2014}; NGC\,5466, and 
NGC\,7006: \citet{jg2010}.

A recent example of this phenomenon are globular clusters associated 
to {\it Gaia Enceladus} (also known as {\it Gaia Sausage}),
a major accretion event that built the stellar halo of the Milky Way
\citep{belokurovetal2018}. Indeed, 9 out of 10 associated globular clusters
have studies of their outer regions and all of them show some of the above mentioned
signatures. For instance,  NGC\,1851, 1904, 2298 and 2808 exhibit tidal tails
\citep{carballobelloetal2018}; extra-tidal features have been found in
NGC\,362 \citep{vanderbekeetal2015}, NGC\,7089 \citep{kuzmaetal2016} and 
NGC\,6779 \citep{pc2019}, while \citet{carballobelloetal2012} mapped the extended 
envelopes of NGC\,1261 and 6864, respectively.

We here exploit the Dark Energy Camera Legacy Survey 
\citep[DECaLS,][]{deyetal2019} in order to address the issue about the existence 
of extra-tidal features around Pal\,13. Section\,2 describes the retrieved data sets, 
while in Section\,3 we deal with the construction of the intrinsic cluster stellar 
density map. The corresponding analysis in the context of the cluster origin and 
kinematics is explained in Section\,4. Finally, Section\,5 summarizes the main 
outcomes of this work.

\section{Data handing}

We downloaded all the information available in the DECaLS.DR8\footnote{http://legacysurvey.org/dr8/} catalogue for an area of 2$\degr$$\times$2$\degr$ centered 
on Pal\,13. The retrieved catalogue contains 306454 sources, for which astrometric and
photometric data are provided homogeneously. As a quality check, we only kept in the
subsequent analysis those sources with morphological model {\sc 'psf'} (stellar point sources),  which resulted to have errors in the PSF $g$ magnitude and $g-r$ color less than 0.04 and 0.07 mag, 
respectively. These upper limits for the photometric errors allow us to deal with a photometry
completeness nearly 100 per cent at $g$ $\le$ 23.0 mag for the outer cluster regions, 
where crowding effects are negligible \citep{deyetal2019}.

\begin{figure*}
\includegraphics[width=\textwidth]{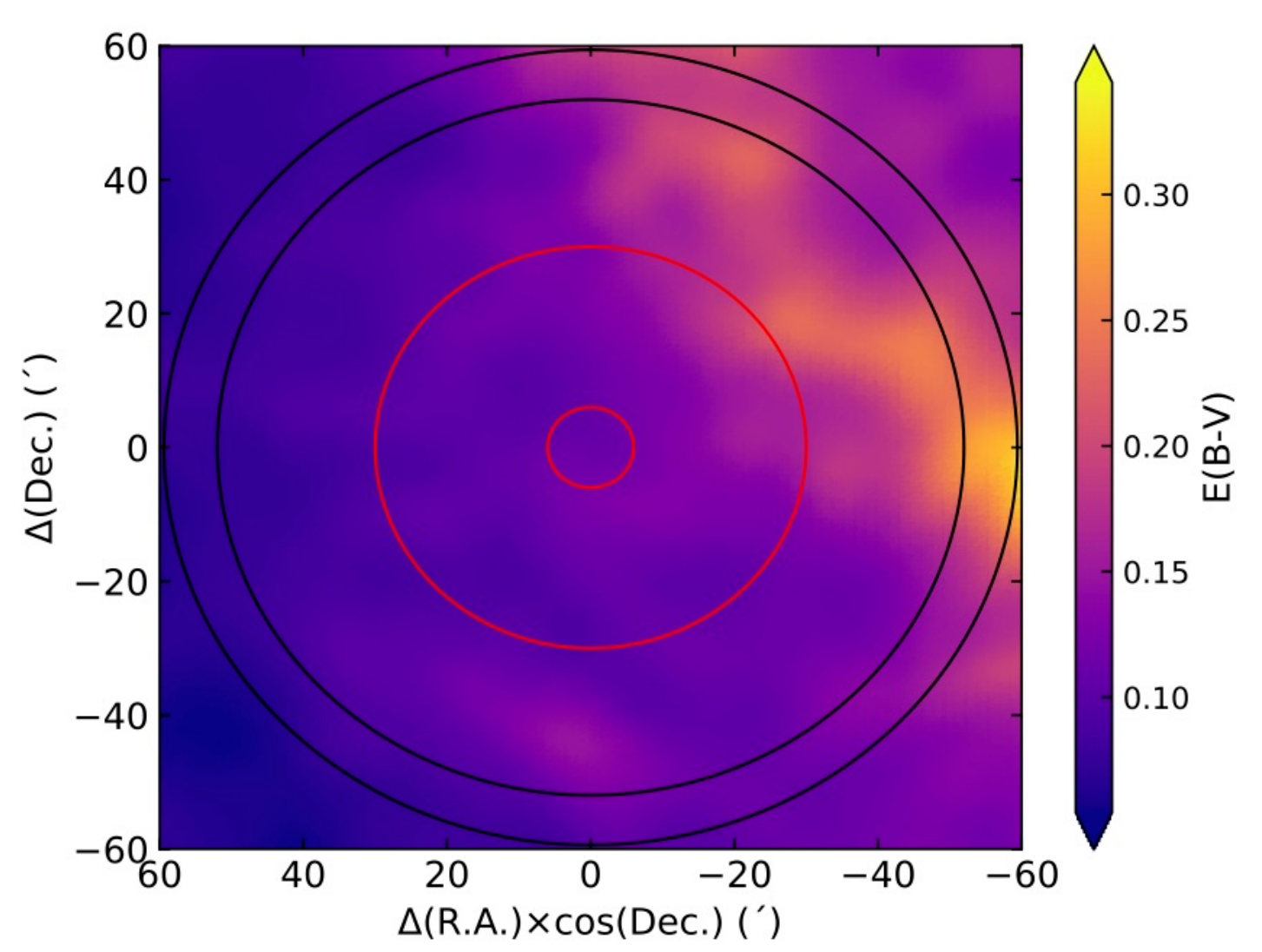}
\caption{Reddening map across the field of Pal\,13. The red circles are of 6$\arcmin$
and 30$\arcmin$ in radius. The outer black annular region is of the same size as
the largest red circular region.}
\label{fig:fig1}
\end{figure*}

The most recent values of the Galactic extinction are also available on the DR8 catalogues,
from which we built the reddening map shown in Fig.~\ref{fig:fig1}. We computed the
average for two circular regions, delineated in the figure by red circles with radii of
6$\arcmin$ and 30$\arcmin$, respectively.  We obtained $<E(B-V)>$ = 0.114$\pm$0.004
mag and 0.117$\pm$0.022 mag for the smallest and largest circles, respectively. These
values show that differential reddening should not mislead our interpretation of the
stellar density distribution around Pal\,13.

\begin{figure*}
\includegraphics[width=\textwidth]{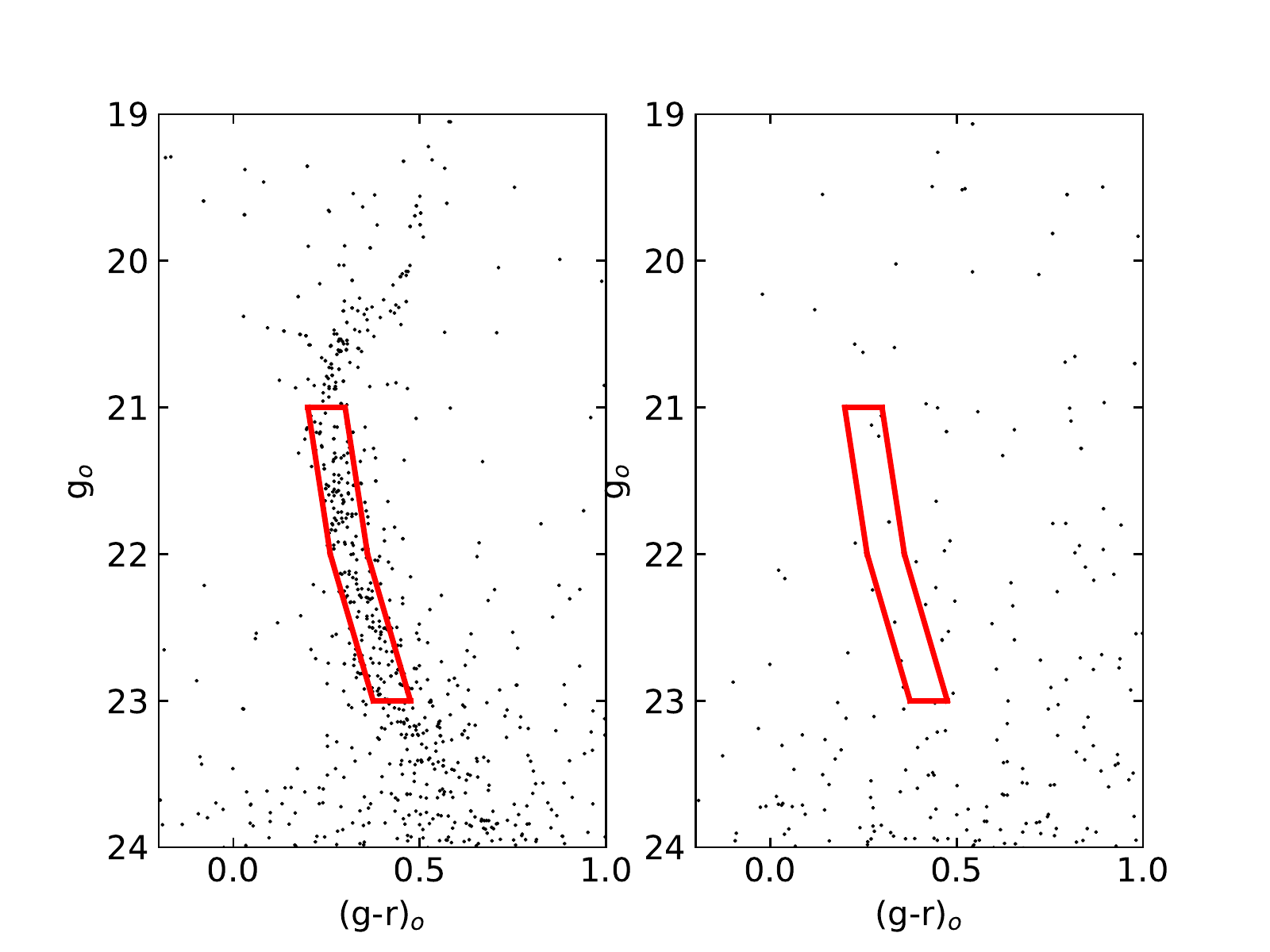}
\caption{CMDs of stars in the field of Pal\,13 ($r<$ 6$\arcmin$; left panel) and in
an annular region of similar area centered on the cluster with outer radius of
60$\arcmin$ (right panel). The region along the cluster MS used to perform star counts is delineated with red contour lines.}

\label{fig:fig2}
\end{figure*}

The intrinsic (reddening corrected)  cluster CMD is depicted in Fig.~\ref{fig:fig2} 
(left panel) for all  the measured stars distributed within the smallest red circle (see 
Fig.~\ref{fig:fig1}).  Note that intrinsic $g_o$ magnitudes and $(g-r)_o$ colors are 
available in the DR8 catalogues. As can be seen,  the main cluster features -including 
some dispersion from field star contamination- are clearly distinguished. For comparison 
purposes, we built a  reference intrinsic star field CMD from stars distributed within an 
annular region located far away from the cluster and with an area equal to that of the 
smallest  red circle (right panel of Fig.~\ref{fig:fig2}). It reveals that the star field
contamination in the cluster CMD should not hamper a reliable tracing  of the
stellar density map in the outer cluster regions.

\section{Stellar density map}

It is well-known that, because of two-body relaxation, the less massive a star,
the outer the cluster regions they can reach. Hence, they are candidates
to cross the Jacobi radius and thus to populate the cluster extra-tidal regions
\citep{carballobelloetal2012}. In previous studies of the external regions of
the globular cluster NGC\,288 \citep{p18a} and NGC\,6779 \citep{pc2019}, we
used a strip along the cluster Main Sequence (MS),  from underneath its MS 
turnoff down to 2 mag, to map the distribution of their stellar populations
beyond their tidal radii. Those stars are low mass enough as to have been
subject of tidal effects, so that we expect that their counterparts in Pal\,13
should also be, particularly because Pal\,13 is less massive than
NGC\,288 and NGC\,6779 \citep{baumgardtetal2019}.

Following the above recipe, we defined the area in the cluster CMD shown
by the red contour in Fig.~\ref{fig:fig2} (left panel) to build the respective
stellar density map. It comprises as many  cluster MS stars as possible and 
minimizes the field star contamination (see right panel of Fig.~\ref{fig:fig2}). 
Nevertheless, we applied a procedure to get rid of field stars that
fall inside the defined strip. The method was devised by \citet{pb12} and
used satisfactorily for cleaning CMDs of star clusters projected towards
crowded star fields  \citep[e.g.,][and references therein]{p17a,p17b,p17c} and affected by differential reddening
\citep[e.g.,][and reference therein]{p2018,petal2018}. It relies on an accurate 
representation of the 
star field CMD in terms of its stellar density, luminosity function and
color distribution. This is done by considering the position of each
field star in the cluster CMD and by subtracting the closest star in the
cluster CMD to that field star. In doing this, we considered the uncertainties
in magnitudes and colors by repeating the procedure hundred of times with
magnitudes and colors varying within their respective errors. 
For the designed MS strip, photometric errors
increase from $\approx$ 0.01 mag up to 0.04 in $g_o$ and from $\approx$ 0.01 mag
up to 0.07 mag in $(g-r)_o$ for the range $g_o$ =  21- 23 mag. 

As for the reference star field, we chose an annular region centered on the cluster,
located relatively far away from it, but  not to far as to lose the star field 
characteristics in the direction toward the cluster. The chosen annulus is meant to embrace
an appropriate collection of any possible star field population and reddening
variation around Pal\,13. The cleaned circular cluster area ($r<$ 30$\arcmin$) and 
the star field annular region are of equal size. The latter is illustrated with
black circles in Fig.~\ref{fig:fig1}.

From the resulting cleaned cluster CMD, we built the stellar density map for
those stars spread within the boundaries of the defined MS strip and located inside
a circle of radius $r$= 30$\arcmin$. We used a kernel density estimator 
(KDE) technique. Particularly, we employed the KDE routine within AstroML \citep{astroml}.
We superimposed a grid of 400$\times$400 squared  cells to the area of
interest and used a range of values for the KDE bandwidth, from 0.3$\arcmin$ up
to 3$\arcmin$ in step of 0.3$\arcmin$, in order to apply the KDE to each generated cell.
 KDE also estimated an optimal bandwidth of 1.5$\arcmin$, which means that
 we resolved structural details larger than $\approx$ 1/7 of $r_t$ = 4.94$\arcmin$
 \citep{baumgardtetal2019}.
  We also estimated the background level using the stars distributed within
 the annular region defined above (black circles in Fig.~\ref{fig:fig1}). We divided such an
 annulus in 16 adjacent sectors of 22.5$\degr$ wide, where we counted the number of stars.
 We rotated such an array of sectors by 11.25$\degr$ and repeated the star counting.
 Finally, we derived the  mean  value in the 32 defined sectors, which
 turned out to be  0.037 stars/arcmin$^2$. As for the standard deviation, 
 we performed 1000 Monte Carlos realizations using the stars located
 beyond 10$\arcmin$ from the cluster center, which were rotated randomly 
 (one different angle for each star) and then the density map recomputed.
 The resulting standard deviation of all the generated density maps 
 turned out to be 0.011 stars/arcmin$^2$.
 
\begin{figure*}
\includegraphics[width=\textwidth]{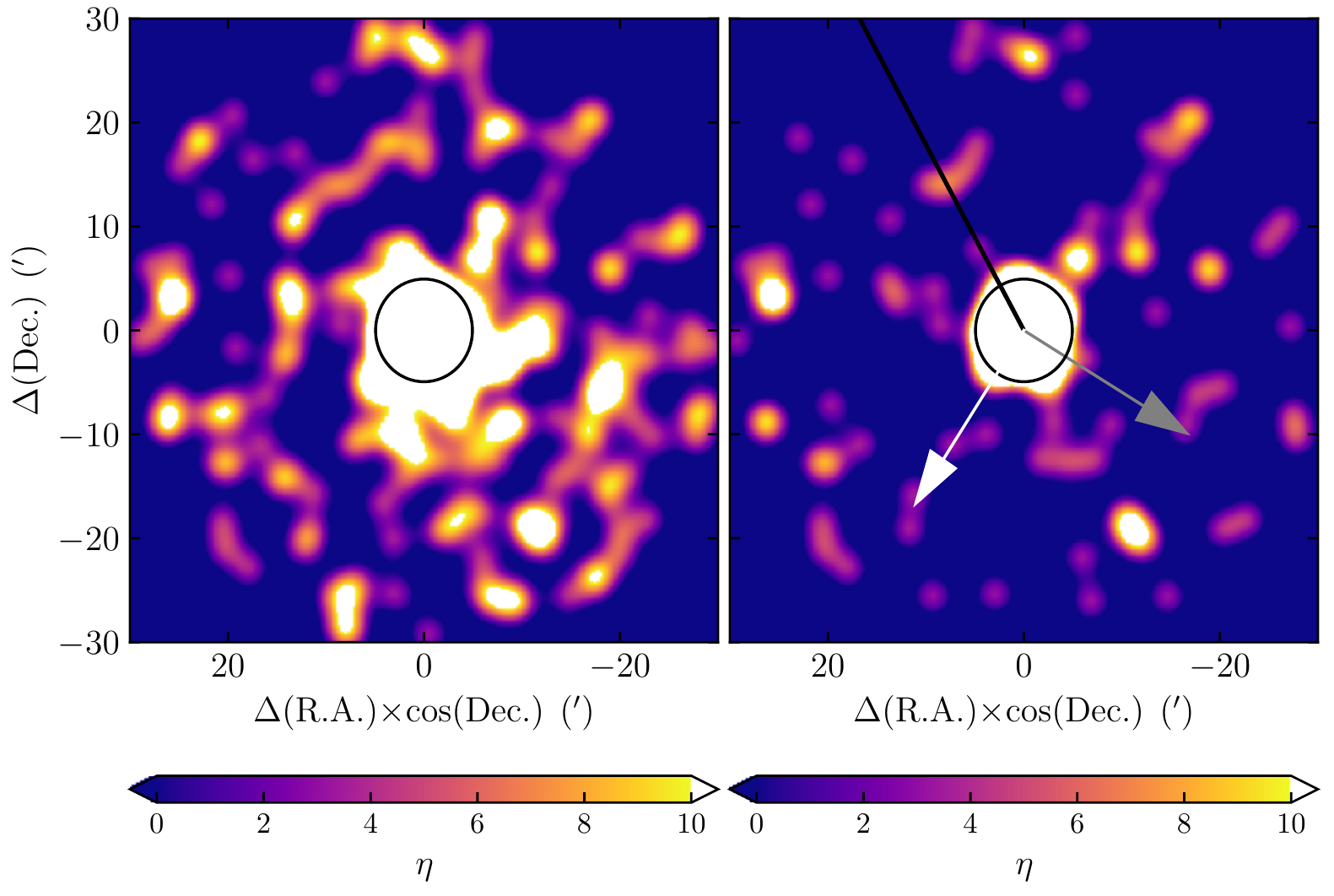}
\caption{Observed (left panel) and field star cleaned (right panel) stellar
density maps, built from stars that occupy the MS strip defined in
Fig.~\ref{fig:fig2}.
The  black circle centered on the cluster indicates the assumed tidal radius
\citet[4.94$\arcmin$][]{baumgardtetal2019}. The different arrows indicate the
directions of the cluster proper motion (gray) and of the Galactic center (white).
The black line represents the cluster orbit computed with GravPot16.}
\label{fig:fig3}
\end{figure*}

\begin{figure}
\includegraphics[width=\columnwidth]{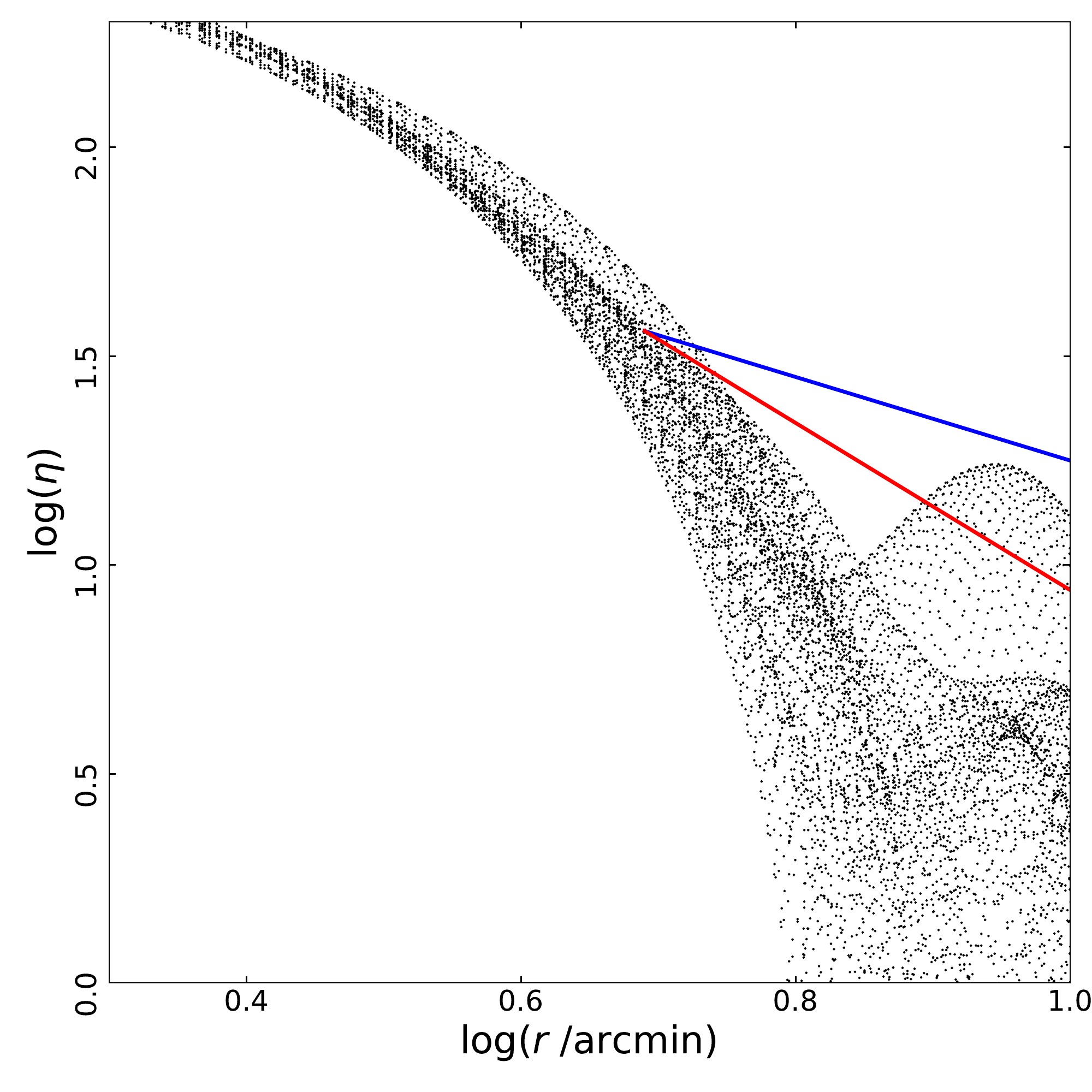}
\caption{ Standard deviations over the mean value in the field as a function of the
distance from the cluster center traced with every point generated from the KDE
technique (see text for details). 
The blue and red lines correspond to a
power law with $\alpha$ = 1 and 2, respectively.}
\label{fig:fig4}
\end{figure}

 The resulting observed and field star cleaned stellar density maps are depicted
 in the left and right panels of Fig.~\ref{fig:fig3}, respectively.  The color scale
 represent the standard deviations over the mean value in the field, i.e., 
 $\eta$ = (signal $-$ 0.037)/0.011. We have painted  white stellar densities higher 
 than  10$\eta$ in order to highlight the cluster less dense stellar   structures. 
 Every point used to generate the field star cleaned density map were 
 also employed to build the cluster stellar radial profile shown in Fig.~\ref{fig:fig4}. 
 
We additionally considered different star field regions distributed beyond 40$\arcmin$
from the cluster centre and applied the same cleaning procedure 
for their respective MS strips using in all the executions the reference star field
as for Pal\,13. We found that the resulting cleaned stellar density maps do not
contain any visible structure above 1$\eta$, which means that
the residuals of the cleaning technique resulted to be negligible.

\begin{figure}
\includegraphics[width=\columnwidth]{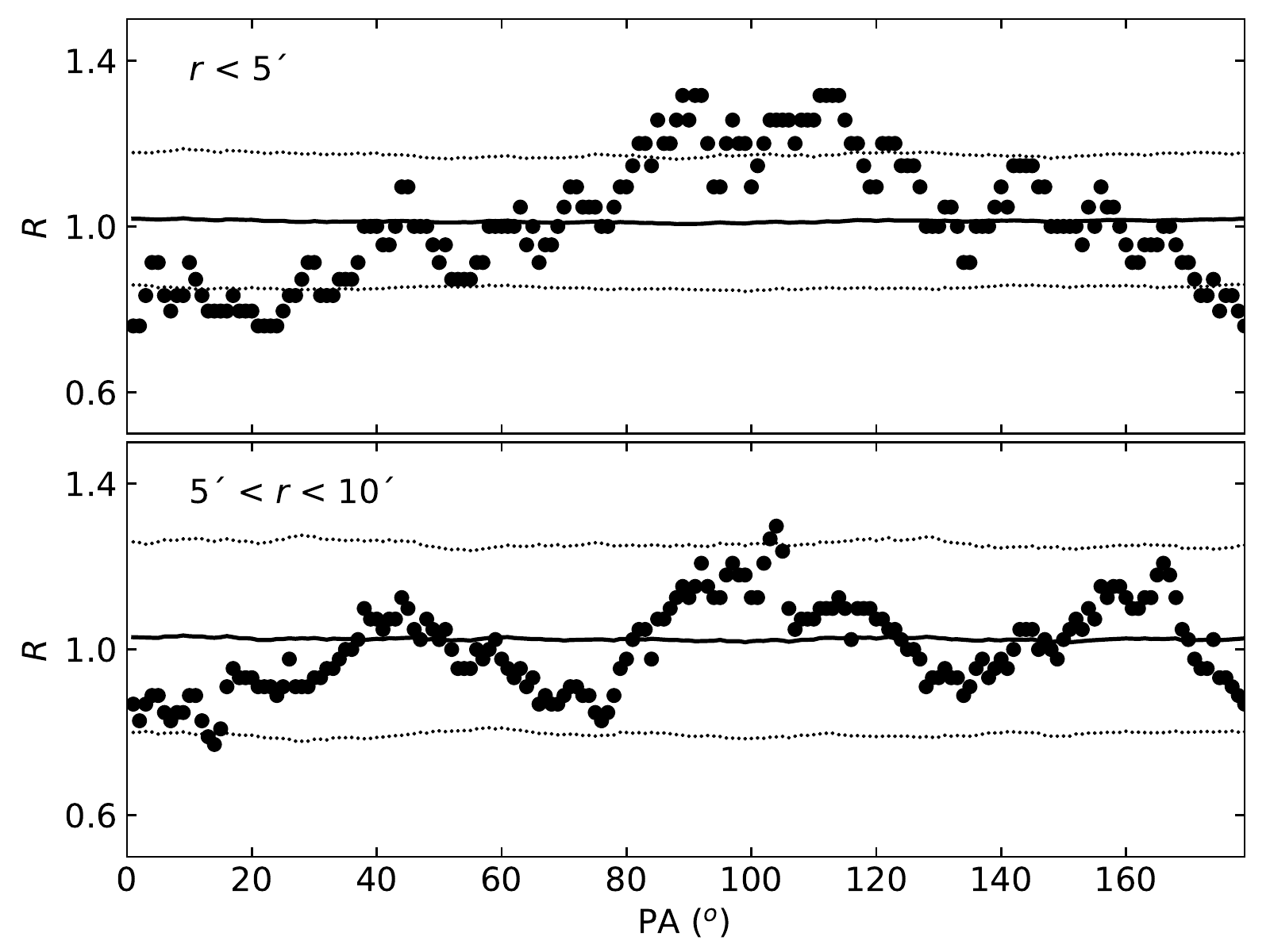}
\caption{$R$ versus PA relationship obtained from star counts in the star
field cleaned MS strip (large filled circles). Solid and dotted lines represent
the resulting mean and dispersion of the Monte Carlo simulations, 
respectively. }
\label{fig:fig5}
\end{figure}

\section{Analysis and discussion}

The observed cluster MS strip stellar density map would seem to suggest
the existence of a concentration of some number of stars from $\approx$ 
5$\arcmin$ out to $\approx$ 10$\arcmin$ from the cluster centre, following 
an azimuthally irregular  pattern (see Fig.~\ref{fig:fig3}, left panel). The lack 
of detection of farther stellar
densities that could be visibly associated to Pal\,13, call our attention on 
the previously derived larger $r_t$ values :  14$\arcmin$$<r_t<$26$\arcmin$:
 \citet{coteetal2002,bradfordetal2011}. To this regard, we recall that field star 
contamination and uncorrected stellar density profiles from incompleteness of 
star counts in  the inner cluster regions because of crowding could lead to fit 
nonphysical motivated \citet{king62} core and tidal radii. 


Here we used a field star cleaning procedure that has turned out to be
effective in getting rid of such a contamination and thus highlighting the
intrinsic extra-tidal features of globular clusters projected on crowded fields or
affected by differential reddening \citep[e.g.][]{pc2019}. Particularly, Pal\,13
would not seem to be projected on either a field affected by a significant differential
reddening (see Section\,2) or on a crowded star field (see Fig.~\ref{fig:fig2},
right panel). From this point of view, the cluster is an easy target for our
purposes, besides the effectiveness of the cleaning method even in more
complicated cluster fields. 
 
The resulting cleaned stellar density map shown in Fig.~\ref{fig:fig3} (right panel) 
exhibits  a more rounded shape than the observed one (Fig.~\ref{fig:fig3}, left panel), 
with some scattered
stellar debris. The figure also shows an excess of stars above 3$\eta$
beyond the adopted limiting radius $r_t$ = 4.94$\arcmin$ ($\equiv$ Jacobi' s radius,
black circle in Fig.~\ref{fig:fig3}), obtained by  \citet{baumgardtetal2019} using eq. 8 in \citet{webbetal2013}. We refer the reader to \citet{piattietal2019b} for a discussion
on the uncertainties of the globular cluster parameters derived by \citet{baumgardtetal2019}
from the integration of their orbital motions. Particularly, they estimated a typical
error of the Jacobi radius of 1.2$\arcmin$ at the Galactocentric distance of
Pal\,13. Fig.~\ref{fig:fig4}  also illustrates the numbers given above. 


We further analyzed the possibility of tidal deformations across the cluster stellar
density map, in the sense that preferential orientation toward the Galactic center
and along the direction of the cluster's orbit are expected in the innermost and
outermost parts, respectively \citep{montuorietal2007}. We followed to recipe applied by
\citet{sollimaetal2011} based on  counts of cluster MS strip stars in alternate
pairs of 90$\degr$ wide circular sectors located at a given distance from the
cluster center and oriented at a position angle (PA) in opposite directions. Then,
we computed the ratio $R(PA) = (N^A_c N^B_f)/(N^A_c N^B_f)$, where $A$ and $B$
are the pair of alternate sectors, and $c$ and $f$ refer to the cluster MS strip
and a CMD field rectangle defined by 21 $<$ $g_o$ (mag) $<$ 23 and 
1.5 < $(g-r)_o$ (mag) < 1.7. In order to assess the statistical significance of our
results, we performed 1000 Monte Carlos realizations using the same number of
measured stars distributed randomly in PA and them obtained the mean and
standard deviations of those independent executions. Fig.~\ref{fig:fig5} depicts the
 resulting curves. As can be seen, there is no noticeable tidal deformations
 across Pal\,13. For completeness purposes we included in Fig.~\ref{fig:fig3}
 (right panel) the directions toward the Galactic center and of the cluster's
 motion. About the globular cluster itself, extra-tidal extensions towards the S-W and
N-E directions are marginally present around the cluster. Pal\,13 has recently undergone
a gravitational shock ($\sim$ 0.6 Gyr according to GravPot16), so that it is very likely
that the extra-tidal extensions visible in Fig.~\ref{fig:fig3} (right panel)
correspond to a very recent disk-shocking, with the extra-tidal material aligned
toward the tidal directions.

Some recent works on the extended structures 
of globular clusters have found that at the outer regions the stellar density is 
$\propto$ $r^{-\alpha}$, with $\alpha$ between 1 and 2 \citep[e.g.,][]{olszewskietal2009,p17c}.
If we assumed a power law decrease of the stellar density for the outermost cluster 
region with slope $\alpha$ = 1 and 2, we would find that Pal\,13 vanishes
 down to 3$\eta$ level at 
$\approx$ 59$\arcmin$ and 17$\arcmin$, respectively. The power law profiles for
$\alpha$ = 1 and 2 are shown in Fig.~\ref{fig:fig4} with blue and red lines,  respectively.
Note that the calculation of $r_t$ involves the cluster
over the course of its orbits, so that it varies from the perigalactic (9.04$\pm$1.74
kpc) up to the apogalactic (67.48$\pm$12.5 kpc) distances between 2.6$\arcmin$ and 
7.2$\arcmin$, respectively \citep{piattietal2019b}. This outcome suggests that its 
present extra-tidal population reaches $\approx$ 1.4$\times$$r_t$.  

Recently, \citet{piatti2019} used the \citet{baumgardtetal2019}'s catalogues to investigate
the kinematics of the Milky Way globular clusters. His results show
that outer globular clusters are prone to have more eccentric orbits (high eccentricity) 
than globular clusters moving in the Milky Way disk, regardless the direction of their motions
(prograde or retrograde orbits). Their orbits also preferentially have large
inclination angles. As far as accreted globular clusters are concerned, they show
radial orbits independently of their position in the Galaxy. Globular clusters whose
orbits have inclination angles $\la$ 50$\degr$ have experienced several disk
crossings compared to those moving along  more circular orbits in the disk at a
similar  Galactocentric distance. Therefore, they have lost more mass \citep{gnedinetal1999,webbetal2014}.   \citet{piattietal2019b} suggested that
the lack of outer clusters rotating in nearly circular orbits (and also low 
inclination angles) could be due to
their accreted origin, while the lack of such  clusters in the inner Milky Way
regions could be due to disruption.

According to \citet{baumgardtetal2019}, Pal\,13 describes a retrograde orbital motion 
($V_{\phi}$ = -73.52  km/s), has an eccentricity of 0.76$\pm$0.05 and and orbital
inclination of 112.26$\degr$$\pm$6.26$\degr$. The ratio of the radial to
total space velocity is 0.86 and the semi-major axis $a$ (average between the
perigalactic and apogalactic distances) is 38.26$\pm$5.50 kpc, respectively.
The ratio of the cluster mass lost due to Milky Way tidal disruption to the
total initial cluster mass computed by \citet{baumgardtetal2019}  is 0.36, 
assuming that the cluster lost half of its initial mass via stellar evolution.
All these features could favor an accreted origin for Pal\,13. Indeed, 
\citet{koposovetal2019} mention this cluster, alongside other 6 ones as
clusters located at $|\phi$$_2| <$ 7$\degr$ of the Orphan stream's great
circle, whose progenitor could be a dwarf galaxy. On the other hand,
 \citet{massarietal2019} using kinematics and chemical
abundance information assigned individual progenitors to most of the known Milky 
Way globular clusters, the {\it Sequoia}  dwarf galaxy being that associated 
one to Pal\,13. {\citet{kc2019} also discuss this issue.}

\citet{kupperetal2011} and \citet{bg2018}  mention that
the cluster is close to its apogalactic distance. \citet{kupperetal2011}  claimed that
because of its position, the cluster should have experienced an expansion that
could have encompassed any extra-tidal structure into the expanded body.
\citet{bg2018} found that cluster tails are more densely packed at 
apogalacticon, so that the cluster should be in the best condition of tails 
(extra-tidal features) observability. From \citet{baumgardtetal2019},
Pal\,13 is currently at a  Galactocentric distance of 25.92 kpc, so that is 1.5 times
farther from its apogalactic distance than from its perigalactic one
(see also the cluster's orbit in \citet{yepezetal2019}). At that position, the
 cluster exhibits a  moderate extended halo of low density. Furthermore,
according to the above estimated $r_t$ values, the extra-tidal corona would also
be observed even at the apogalactic distance.

\begin{figure*}
\includegraphics[width=\textwidth]{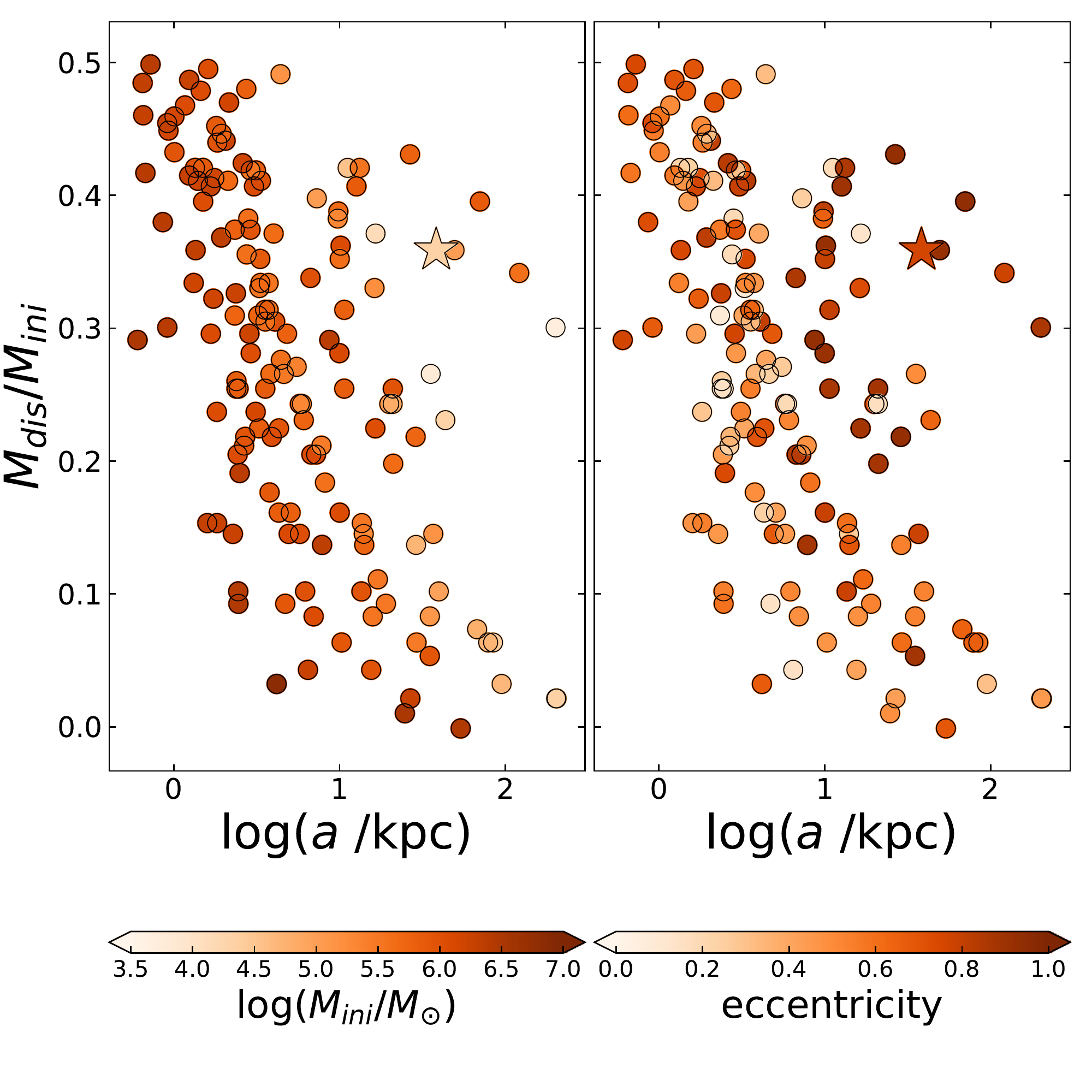}
\caption{Relationship between the ratio of the cluster mass lost
by tidal effects to the initial cluster mass as a function of the
cluster  orbit semi-major axis for the entire sample of Milky Way globular clusters in
the \citet{baumgardtetal2019}'s catalogues. colored symbols are according to
the respective horizontal color bars. Pal\,13 is represented by a large starred
symbol.}
\label{fig:fig6}
\end{figure*}

The presence of extra-tidal features in Pal\,13  could be alternatively related
with the  relatively
large estimated ratio of mass lost by tidal disrution (0.36). \citet{hamrenetal2013}
showed that low-mass globular cluster may have lost a considerable
amount of mass as compared to those more massive clusters. Fig.~\ref{fig:fig6}
(left panel) shows the ratio of the cluster mass disrupted by tidal effects to the
total cluster mass as a function of the semi-major axis for the entire sample
of globular clusters in \citet{baumgardtetal2019}. Points have been
colored according to the initial mass as indicated by the horizontal color bar.
Pal\,13 is shown with a large starred symbol. As can be seen, there is not
a clear correlation between the mass lost and the initial cluster mass for
clusters at a similar semi-major axis as Pal\,13, as
suggested by \citet{hamrenetal2013}. However, if we reproduce that
plot using the eccentricity as the colored scale variable, a clearer correlation
arises, in the sense that the higher the orbital eccentricity of a 
globular cluster with similar semi-major axis as Pal\,13, the higher
the mass lost by tidal disruption. This outcome is in the line of the recent
\citet{piattietal2019b}'s results. Their figure 1 shows that the cluster
eccentricity plays a role in the amount of cluster mass lost by tidal
disruption, because of the more numerous disk crossing experienced 
respect to clusters with more rounded orbital motions. Likewise,
their figure 4 reflects that present-day cluster limiting radii, for clusters
with similar semi-major axes, are smaller for those that have lost
larger amount of mass due to tidal effects, which we corroborate
for Pal\,13. These results are fully compatible with Pal\,13 being
relatively compact and at the same time exhibits extra-tidal
features.

\section{Conclusions}

In this work we took advantage of the DECaLS DR8 data sets to address the
issue about the existence of extra-tidal structures around Pal\,13, one of the relatively 
compact and less massive known Milky Way globular clusters. Precisely
because of its size, some previous works have concluded that the cluster does not
extent beyond its Jacobi radius, and  that the farthest low-mass members
are quite distant from that boundary. Such a result has not been conclusive,
since complementary works based on {\it Gaia} DR2 proper motions or surface
stellar density profiles have found evidence about the cluster is being tidally stripped. 
Additionally, there has been no consensus about its actual dimension, as judged by the 
wide range of values derived for its tidal radius.

As far as we are aware, the Panoramic Survey Telescope and Rapid response System
\citep[Pan-STARRS PS1,][]{chambersetal2016} and DECaLS are the only present 
publicly available astrometric and photometric surveys covering a wide area around 
Pal\,13, making them valuable tools for studying the cluster extended structures. We 
decided to use DELCaLS DR8 because of its comparative better performance 
in terms of limiting magnitude, photometric errors and completeness. Such a data set 
was employed to build a stellar density map of stars distributed along the cluster MS,
 down to 2 mag below the MS turnoff.



At first glance, the cluster stellar density map shows nearly smooth 
contours around the outer cluster regions, that seem to reach out to $\approx$ 1.6 its limiting
radius. Here, we adopted the most recent value of $r_t$ = 4.94$\arcmin$
derived by \citet{baumgardtetal2019} using line-of-sight velocity and proper motion 
velocity dispersion profiles fitted from a grid of dedicated $N$-body simulations.
This outcome favours the presence of stars escaping the cluster, a phenomenon
frequently seen in globular clusters that have crossed the Milky Way disc
 a large number of times \citep{piattietal2019b}. Indeed, the high 
eccentricity and large inclination angle of the Pal\,13's orbital motion help us to 
speculate on that possibility. Furthermore, because of its retrograde direction
of motion, Pal\,13 has been  suggested to have an accreted origin, particularly
associated to the {\it Sequoia} progenitor when its age and overall metallicity
are also considered \citep{massarietal2019}. Others six globular clusters
could have been formed in the same dwarf galaxy, and four of them with
studies of their external regions also have extra-tidal features.

\begin{acknowledgements}
We thank the referee for the thorough reading of the manuscript and
timely suggestions to improve it. 

We thank Eugene Vasiliev for letting us know about the DECaLS public data sets for
Pal\,13.

The Legacy Surveys consist of three individual and complementary projects: the Dark Energy Camera Legacy Survey (DECaLS; NOAO Proposal ID \# 2014B-0404; PIs: David Schlegel and Arjun Dey), the Beijing-Arizona Sky Survey (BASS; NOAO Proposal ID \# 2015A-0801; PIs: Zhou Xu and Xiaohui Fan), and the Mayall z-band Legacy Survey (MzLS; NOAO Proposal ID \# 2016A-0453; PI: Arjun Dey). DECaLS, BASS and MzLS together include data obtained, respectively, at the Blanco telescope, Cerro Tololo Inter-American Observatory, National Optical Astronomy Observatory (NOAO); the Bok telescope, Steward Observatory, University of Arizona; and the Mayall telescope, Kitt Peak National Observatory, NOAO. The Legacy Surveys project is honored to be permitted to conduct astronomical research on Iolkam Du'ag (Kitt Peak), a mountain with particular significance to the Tohono O'odham Nation.

NOAO is operated by the Association of Universities for Research in Astronomy (AURA) under a cooperative agreement with the National Science Foundation.

This project used data obtained with the Dark Energy Camera (DECam), which was constructed by the Dark Energy Survey (DES) collaboration. Funding for the DES Projects has been provided by the U.S. Department of Energy, the U.S. National Science Foundation, the Ministry of Science and Education of Spain, the Science and Technology Facilities Council of the United Kingdom, the Higher Education Funding Council for England, the National Center for Supercomputing Applications at the University of Illinois at Urbana-Champaign, the Kavli Institute of Cosmological Physics at the University of Chicago, Center for Cosmology and Astro-Particle Physics at the Ohio State University, the Mitchell Institute for Fundamental Physics and Astronomy at Texas A\&M University, Financiadora de Estudos e Projetos, Fundacao Carlos Chagas Filho de Amparo, Financiadora de Estudos e Projetos, Fundacao Carlos Chagas Filho de Amparo a Pesquisa do Estado do Rio de Janeiro, Conselho Nacional de Desenvolvimento Cientifico e Tecnologico and the Ministerio da Ciencia, Tecnologia e Inovacao, the Deutsche Forschungsgemeinschaft and the Collaborating Institutions in the Dark Energy Survey. The Collaborating Institutions are Argonne National Laboratory, the University of California at Santa Cruz, the University of Cambridge, Centro de Investigaciones Energeticas, Medioambientales y Tecnologicas-Madrid, the University of Chicago, University College London, the DES-Brazil Consortium, the University of Edinburgh, the Eidgenossische Technische Hochschule (ETH) Zurich, Fermi National Accelerator Laboratory, the University of Illinois at Urbana-Champaign, the Institut de Ciencies de l'Espai (IEEC/CSIC), the Institut de Fisica d'Altes Energies, Lawrence Berkeley National Laboratory, the Ludwig-Maximilians Universitat Munchen and the associated Excellence Cluster Universe, the University of Michigan, the National Optical Astronomy Observatory, the University of Nottingham, the Ohio State University, the University of Pennsylvania, the University of Portsmouth, SLAC National Accelerator Laboratory, Stanford University, the University of Sussex, and Texas A\&M University.

BASS is a key project of the Telescope Access Program (TAP), which has been funded by the National Astronomical Observatories of China, the Chinese Academy of Sciences (the Strategic Priority Research Program "The Emergence of Cosmological Structures" Grant \# XDB09000000), and the Special Fund for Astronomy from the Ministry of Finance. The BASS is also supported by the External Cooperation Program of Chinese Academy of Sciences (Grant \# 114A11KYSB20160057), and Chinese National Natural Science Foundation (Grant \# 11433005).

The Legacy Survey team makes use of data products from the Near-Earth Object Wide-field Infrared Survey Explorer (NEOWISE), which is a project of the Jet Propulsion Laboratory/California Institute of Technology. NEOWISE is funded by the National Aeronautics and Space Administration.

The Legacy Surveys imaging of the DESI footprint is supported by the Director, Office of Science, Office of High Energy Physics of the U.S. Department of Energy under Contract No. DE-AC02-05CH1123, by the National Energy Research Scientific Computing Center, a DOE Office of Science User Facility under the same contract; and by the U.S. National Science Foundation, Division of Astronomical Sciences under Contract No. AST-0950945 to NOAO.

A.E.P. acknowledge support from the Ministerio de Ciencia, Tecnolog\'{\i}a e Innovaci\'on Productiva (MINCyT) through grant PICT-201-0030. J.G.F-T is supported by FONDECYT No. 3180210 and Becas
Iberoam\'erica Investigador 2019, banco Santarder Chile. 
\end{acknowledgements}


\begin{thebibliography}{54}
\expandafter\ifx\csname natexlab\endcsname\relax\def\natexlab#1{#1}\fi

\bibitem[{{Balbinot} \& {Gieles}(2018)}]{bg2018}
{Balbinot}, E. \& {Gieles}, M. 2018, \mnras, 474, 2479

\bibitem[{{Barb{\'a}} {et~al.}(2019){Barb{\'a}}, {Minniti}, {Geisler},
  {Alonso-Garc{\'\i}a}, {Hempel}, {Monachesi}, {Arias}, \&
  {G{\'o}mez}}]{barbaetal2019}
{Barb{\'a}}, R.~H., {Minniti}, D., {Geisler}, D., {et~al.} 2019, \apjl, 870,
  L24

\bibitem[{{Baumgardt} {et~al.}(2019){Baumgardt}, {Hilker}, {Sollima}, \&
  {Bellini}}]{baumgardtetal2019}
{Baumgardt}, H., {Hilker}, M., {Sollima}, A., \& {Bellini}, A. 2019, \mnras,
  482, 5138

\bibitem[{{Belokurov} {et~al.}(2018){Belokurov}, {Erkal}, {Evans}, {Koposov},
  \& {Deason}}]{belokurovetal2018}
{Belokurov}, V., {Erkal}, D., {Evans}, N.~W., {Koposov}, S.~E., \& {Deason},
  A.~J. 2018, \mnras, 478, 611

\bibitem[{{Belokurov} {et~al.}(2006){Belokurov}, {Evans}, {Irwin}, {Hewett}, \&
  {Wilkinson}}]{belokurovetal2006}
{Belokurov}, V., {Evans}, N.~W., {Irwin}, M.~J., {Hewett}, P.~C., \&
  {Wilkinson}, M.~I. 2006, \apjl, 637, L29

\bibitem[{{Bradford} {et~al.}(2011){Bradford}, {Geha}, {Mu{\~n}oz}, {Santana},
  {Simon}, {C{\^o}t{\'e}}, {Stetson}, {Kirby}, \&
  {Djorgovski}}]{bradfordetal2011}
{Bradford}, J.~D., {Geha}, M., {Mu{\~n}oz}, R.~R., {et~al.} 2011, \apj, 743,
  167

\bibitem[{{Carballo-Bello} {et~al.}(2012){Carballo-Bello}, {Gieles}, {Sollima},
  {Koposov}, {Mart{\'{\i}}nez-Delgado}, \&
  {Pe{\~n}arrubia}}]{carballobelloetal2012}
{Carballo-Bello}, J.~A., {Gieles}, M., {Sollima}, A., {et~al.} 2012, \mnras,
  419, 14

\bibitem[{{Carballo-Bello} {et~al.}(2018){Carballo-Bello},
  {Mart{\'{\i}}nez-Delgado}, {Navarrete}, {Catelan}, {Mu{\~n}oz}, {Antoja}, \&
  {Sollima}}]{carballobelloetal2018}
{Carballo-Bello}, J.~A., {Mart{\'{\i}}nez-Delgado}, D., {Navarrete}, C.,
  {et~al.} 2018, \mnras, 474, 683

\bibitem[{{Chambers} {et~al.}(2016){Chambers}, {Magnier}, {Metcalfe},
  {Flewelling}, {Huber}, {Waters}, {Denneau}, {Draper}, {Farrow}, {Finkbeiner},
  {Holmberg}, {Koppenhoefer}, {Price}, {Saglia}, {Schlafly}, {Smartt},
  {Sweeney}, {Wainscoat}, {Burgett}, {Grav}, {Heasley}, {Hodapp}, {Jedicke},
  {Kaiser}, {Kudritzki}, {Luppino}, {Lupton}, {Monet}, {Morgan}, {Onaka},
  {Stubbs}, {Tonry}, {Banados}, {Bell}, {Bender}, {Bernard}, {Botticella},
  {Casertano}, {Chastel}, {Chen}, {Chen}, {Cole}, {Deacon}, {Frenk},
  {Fitzsimmons}, {Gezari}, {Goessl}, {Goggia}, {Goldman}, {Grebel}, {Hambly},
  {Hasinger}, {Heavens}, {Heckman}, {Henderson}, {Henning}, {Holman}, {Hopp},
  {Ip}, {Isani}, {Keyes}, {Koekemoer}, {Kotak}, {Long}, {Lucey}, {Liu},
  {Martin}, {McLean}, {Morganson}, {Murphy}, {Nieto-Santisteban}, {Norberg},
  {Peacock}, {Pier}, {Postman}, {Primak}, {Rae}, {Rest}, {Riess}, {Riffeser},
  {Rix}, {Roser}, {Schilbach}, {Schultz}, {Scolnic}, {Szalay}, {Seitz},
  {Shiao}, {Small}, {Smith}, {Soderblom}, {Taylor}, {Thakar}, {Thiel},
  {Thilker}, {Urata}, {Valenti}, {Walter}, {Watters}, {Werner}, {White},
  {Wood-Vasey}, \& {Wyse}}]{chambersetal2016}
{Chambers}, K.~C., {Magnier}, E.~A., {Metcalfe}, N., {et~al.} 2016, ArXiv
  e-prints [\eprint[arXiv]{1612.05560}]

\bibitem[{{Correnti} {et~al.}(2011){Correnti}, {Bellazzini}, {Dalessandro},
  {Mucciarelli}, {Monaco}, \& {Catelan}}]{correntietal2011}
{Correnti}, M., {Bellazzini}, M., {Dalessandro}, E., {et~al.} 2011, \mnras,
  417, 2411

\bibitem[{{C{\^o}t{\'e}} {et~al.}(2002){C{\^o}t{\'e}}, {Djorgovski}, {Meylan},
  {Castro}, \& {McCarthy}}]{coteetal2002}
{C{\^o}t{\'e}}, P., {Djorgovski}, S.~G., {Meylan}, G., {Castro}, S., \&
  {McCarthy}, J.~K. 2002, \apj, 574, 783

\bibitem[{{Dey} {et~al.}(2019){Dey}, {Schlegel}, {Lang}, {Blum}, {Burleigh},
  {Fan}, {Findlay}, {Finkbeiner}, {Herrera}, {Juneau}, {Landriau}, {Levi},
  {McGreer}, {Meisner}, {Myers}, {Moustakas}, {Nugent}, {Patej}, {Schlafly},
  {Walker}, {Valdes}, {Weaver}, {Y{\`e}che}, {Zou}, {Zhou}, {Abareshi},
  {Abbott}, {Abolfathi}, {Aguilera}, {Alam}, {Allen}, {Alvarez}, {Annis},
  {Ansarinejad}, {Aubert}, {Beechert}, {Bell}, {BenZvi}, {Beutler}, {Bielby},
  {Bolton}, {Brice{\~n}o}, {Buckley-Geer}, {Butler}, {Calamida}, {Carlberg},
  {Carter}, {Casas}, {Castander}, {Choi}, {Comparat}, {Cukanovaite}, {Delubac},
  {DeVries}, {Dey}, {Dhungana}, {Dickinson}, {Ding}, {Donaldson}, {Duan},
  {Duckworth}, {Eftekharzadeh}, {Eisenstein}, {Etourneau}, {Fagrelius},
  {Farihi}, {Fitzpatrick}, {Font-Ribera}, {Fulmer}, {G{\"a}nsicke},
  {Gaztanaga}, {George}, {Gerdes}, {Gontcho}, {Gorgoni}, {Green}, {Guy},
  {Harmer}, {Hernand ez}, {Honscheid}, {Huang}, {James}, {Jannuzi}, {Jiang},
  {Joyce}, {Karcher}, {Karkar}, {Kehoe}, {Kneib}, {Kueter-Young}, {Lan},
  {Lauer}, {Le Guillou}, {Le Van Suu}, {Lee}, {Lesser}, {Perreault Levasseur},
  {Li}, {Mann}, {Marshall}, {Mart{\'\i}nez-V{\'a}zquez}, {Martini}, {du Mas des
  Bourboux}, {McManus}, {Meier}, {M{\'e}nard}, {Metcalfe},
  {Mu{\~n}oz-Guti{\'e}rrez}, {Najita}, {Napier}, {Narayan}, {Newman}, {Nie},
  {Nord}, {Norman}, {Olsen}, {Paat}, {Palanque-Delabrouille}, {Peng},
  {Poppett}, {Poremba}, {Prakash}, {Rabinowitz}, {Raichoor}, {Rezaie},
  {Robertson}, {Roe}, {Ross}, {Ross}, {Rudnick}, {Safonova}, {Saha},
  {S{\'a}nchez}, {Savary}, {Schweiker}, {Scott}, {Seo}, {Shan}, {Silva},
  {Slepian}, {Soto}, {Sprayberry}, {Staten}, {Stillman}, {Stupak}, {Summers},
  {Sien Tie}, {Tirado}, {Vargas-Maga{\~n}a}, {Vivas}, {Wechsler}, {Williams},
  {Yang}, {Yang}, {Yapici}, {Zaritsky}, {Zenteno}, {Zhang}, {Zhang}, {Zhou}, \&
  {Zhou}}]{deyetal2019}
{Dey}, A., {Schlegel}, D.~J., {Lang}, D., {et~al.} 2019, \aj, 157, 168

\bibitem[{{Fern{\'a}ndez-Trincado} {et~al.}(2019){Fern{\'a}ndez-Trincado},
  {Beers}, {Placco}, {Moreno}, {Alves-Brito}, {Minniti}, {Tang},
  {P{\'e}rez-Villegas}, {Reyl{\'e}}, {Robin}, \& {Villanova}}]{fetal2019a}
{Fern{\'a}ndez-Trincado}, J.~G., {Beers}, T.~C., {Placco}, V.~M., {et~al.}
  2019, \apjl, 886, L8

\bibitem[{{Gnedin} {et~al.}(1999){Gnedin}, {Lee}, \&
  {Ostriker}}]{gnedinetal1999}
{Gnedin}, O.~Y., {Lee}, H.~M., \& {Ostriker}, J.~P. 1999, \apj, 522, 935

\bibitem[{{Grillmair}(2019)}]{g2019}
{Grillmair}, C.~J. 2019, arXiv e-prints, arXiv:1909.05927

\bibitem[{{Hamren} {et~al.}(2013){Hamren}, {Smith}, {Guhathakurta}, {Dolphin},
  {Weisz}, {Rajan}, \& {Grillmair}}]{hamrenetal2013}
{Hamren}, K.~M., {Smith}, G.~H., {Guhathakurta}, P., {et~al.} 2013, \aj, 146,
  116

\bibitem[{{Harris}(1996)}]{harris1996}
{Harris}, W.~E. 1996, \aj, 112, 1487

\bibitem[{{Ibata} {et~al.}(2019){Ibata}, {Bellazzini}, {Malhan}, {Martin}, \&
  {Bianchini}}]{ibataetal2019}
{Ibata}, R.~A., {Bellazzini}, M., {Malhan}, K., {Martin}, N., \& {Bianchini},
  P. 2019, Nature Astronomy, 3, 667

\bibitem[{{Jordi} \& {Grebel}(2010)}]{jg2010}
{Jordi}, K. \& {Grebel}, E.~K. 2010, \aap, 522, A71

\bibitem[{{King}(1962)}]{king62}
{King}, I. 1962, \aj, 67, 471

\bibitem[{{Koch} \& {C{\^o}t{\'e}}(2019)}]{kc2019}
{Koch}, A. \& {C{\^o}t{\'e}}, P. 2019, arXiv e-prints, arXiv:1910.13506

\bibitem[{{Koposov} {et~al.}(2019){Koposov}, {Belokurov}, {Li}, {Mateu},
  {Erkal}, {Grillmair}, {Hendel}, {Price-Whelan}, {Laporte}, {Hawkins}, {Sohn},
  {del Pino}, {Evans}, {Slater}, {Kallivayalil}, {Navarro}, \& {Orphan Aspen
  Treasury Collaboration}}]{koposovetal2019}
{Koposov}, S.~E., {Belokurov}, V., {Li}, T.~S., {et~al.} 2019, \mnras, 485,
  4726

\bibitem[{{Kunder} {et~al.}(2014){Kunder}, {Bono}, {Piffl}, {Steinmetz},
  {Grebel}, {Anguiano}, {Freeman}, {Kordopatis}, {Zwitter}, {Scholz}, {Gibson},
  {Bland-Hawthorn}, {Seabroke}, {Boeche}, {Siebert}, {Wyse}, {Bienaym{\'e}},
  {Navarro}, {Siviero}, {Minchev}, {Parker}, {Reid}, {Gilmore}, {Munari}, \&
  {Helmi}}]{kunderetal2014}
{Kunder}, A., {Bono}, G., {Piffl}, T., {et~al.} 2014, \aap, 572, A30

\bibitem[{{Kundu} {et~al.}(2019){Kundu}, {Minniti}, \& {Singh}}]{kunduetal2019}
{Kundu}, R., {Minniti}, D., \& {Singh}, H.~P. 2019, \mnras, 483, 1737

\bibitem[{{K{\"u}pper} {et~al.}(2010){K{\"u}pper}, {Kroupa}, {Baumgardt}, \&
  {Heggie}}]{kupperetal2010}
{K{\"u}pper}, A.~H.~W., {Kroupa}, P., {Baumgardt}, H., \& {Heggie}, D.~C. 2010,
  \mnras, 401, 105

\bibitem[{{K{\"u}pper} {et~al.}(2011){K{\"u}pper}, {Mieske}, \&
  {Kroupa}}]{kupperetal2011}
{K{\"u}pper}, A. H.~W., {Mieske}, S., \& {Kroupa}, P. 2011, \mnras, 413, 863

\bibitem[{{Kuzma} {et~al.}(2016){Kuzma}, {Da Costa}, {Mackey}, \&
  {Roderick}}]{kuzmaetal2016}
{Kuzma}, P.~B., {Da Costa}, G.~S., {Mackey}, A.~D., \& {Roderick}, T.~A. 2016,
  \mnras, 461, 3639

\bibitem[{{Massari} {et~al.}(2019){Massari}, {Koppelman}, \&
  {Helmi}}]{massarietal2019}
{Massari}, D., {Koppelman}, H.~H., \& {Helmi}, A. 2019, \aap, 630, L4

\bibitem[{{Montuori} {et~al.}(2007){Montuori}, {Capuzzo-Dolcetta}, {Di Matteo},
  {Lepinette}, \& {Miocchi}}]{montuorietal2007}
{Montuori}, M., {Capuzzo-Dolcetta}, R., {Di Matteo}, P., {Lepinette}, A., \&
  {Miocchi}, P. 2007, \apj, 659, 1212

\bibitem[{{Myeong} {et~al.}(2017){Myeong}, {Jerjen}, {Mackey}, \& {Da
  Costa}}]{myeongetal2017}
{Myeong}, G.~C., {Jerjen}, H., {Mackey}, D., \& {Da Costa}, G.~S. 2017, \apjl,
  840, L25

\bibitem[{{Myeong} {et~al.}(2019){Myeong}, {Vasiliev}, {Iorio}, {Evans}, \&
  {Belokurov}}]{myeongetal2019}
{Myeong}, G.~C., {Vasiliev}, E., {Iorio}, G., {Evans}, N.~W., \& {Belokurov},
  V. 2019, \mnras, 488, 1235

\bibitem[{{Navarrete} {et~al.}(2017){Navarrete}, {Belokurov}, \&
  {Koposov}}]{naverreteetal2017}
{Navarrete}, C., {Belokurov}, V., \& {Koposov}, S.~E. 2017, \apjl, 841, L23

\bibitem[{{Niederste-Ostholt} {et~al.}(2010){Niederste-Ostholt}, {Belokurov},
  {Evans}, {Koposov}, {Gieles}, \& {Irwin}}]{noetal2010}
{Niederste-Ostholt}, M., {Belokurov}, V., {Evans}, N.~W., {et~al.} 2010,
  \mnras, 408, L66

\bibitem[{{Odenkirchen} {et~al.}(2003){Odenkirchen}, {Grebel}, {Dehnen}, {Rix},
  {Yanny}, {Newberg}, {Rockosi}, {Mart{\'{\i}}nez-Delgado}, {Brinkmann}, \&
  {Pier}}]{odenetal2003}
{Odenkirchen}, M., {Grebel}, E.~K., {Dehnen}, W., {et~al.} 2003, \aj, 126, 2385

\bibitem[{{Olszewski} {et~al.}(2009){Olszewski}, {Saha}, {Knezek},
  {Subramaniam}, {de Boer}, \& {Seitzer}}]{olszewskietal2009}
{Olszewski}, E.~W., {Saha}, A., {Knezek}, P., {et~al.} 2009, \aj, 138, 1570

\bibitem[{{Piatti}(2017{\natexlab{a}})}]{p17c}
{Piatti}, A.~E. 2017{\natexlab{a}}, \apjl, 846, L10

\bibitem[{{Piatti}(2017{\natexlab{b}})}]{p17a}
{Piatti}, A.~E. 2017{\natexlab{b}}, \apjl, 834, L14

\bibitem[{{Piatti}(2017{\natexlab{c}})}]{p17b}
{Piatti}, A.~E. 2017{\natexlab{c}}, \mnras, 465, 2748

\bibitem[{{Piatti}(2018{\natexlab{a}})}]{p18a}
{Piatti}, A.~E. 2018{\natexlab{a}}, \mnras, 473, 492

\bibitem[{{Piatti}(2018{\natexlab{b}})}]{p2018}
{Piatti}, A.~E. 2018{\natexlab{b}}, \mnras, 477, 2164

\bibitem[{{Piatti}(2019)}]{piatti2019}
{Piatti}, A.~E. 2019, \apj, 882, 98

\bibitem[{{Piatti} \& {Bica}(2012)}]{pb12}
{Piatti}, A.~E. \& {Bica}, E. 2012, \mnras, 425, 3085

\bibitem[{{Piatti} \& {Carballo-Bello}(2019)}]{pc2019}
{Piatti}, A.~E. \& {Carballo-Bello}, J.~A. 2019, \mnras, 485, 1029

\bibitem[{{Piatti} {et~al.}(2018){Piatti}, {Cole}, \& {Emptage}}]{petal2018}
{Piatti}, A.~E., {Cole}, A.~A., \& {Emptage}, B. 2018, \mnras, 473, 105

\bibitem[{{Piatti} {et~al.}(2019){Piatti}, {Webb}, \&
  {Carlberg}}]{piattietal2019b}
{Piatti}, A.~E., {Webb}, J.~J., \& {Carlberg}, R.~G. 2019, \mnras, 489, 4367

\bibitem[{{Shipp} {et~al.}(2018){Shipp}, {Drlica-Wagner}, {Balbinot},
  {Ferguson}, {Erkal}, {Li}, {Bechtol}, {Belokurov}, {Buncher}, {Carollo},
  {Carrasco Kind}, {Kuehn}, {Marshall}, {Pace}, {Rykoff}, {Sevilla-Noarbe},
  {Sheldon}, {Strigari}, {Vivas}, {Yanny}, {Zenteno}, {Abbott}, {Abdalla},
  {Allam}, {Avila}, {Bertin}, {Brooks}, {Burke}, {Carretero}, {Castander},
  {Cawthon}, {Crocce}, {Cunha}, {D'Andrea}, {da Costa}, {Davis}, {De Vicente},
  {Desai}, {Diehl}, {Doel}, {Evrard}, {Flaugher}, {Fosalba}, {Frieman},
  {Garc{\'\i}a-Bellido}, {Gaztanaga}, {Gerdes}, {Gruen}, {Gruendl}, {Gschwend},
  {Gutierrez}, {Hartley}, {Honscheid}, {Hoyle}, {James}, {Johnson}, {Krause},
  {Kuropatkin}, {Lahav}, {Lin}, {Maia}, {March}, {Martini}, {Menanteau},
  {Miller}, {Miquel}, {Nichol}, {Plazas}, {Romer}, {Sako}, {Sanchez},
  {Santiago}, {Scarpine}, {Schindler}, {Schubnell}, {Smith}, {Smith},
  {Sobreira}, {Suchyta}, {Swanson}, {Tarle}, {Thomas}, {Tucker}, {Walker},
  {Wechsler}, \& {DES Collaboration}}]{shippetal2018}
{Shipp}, N., {Drlica-Wagner}, A., {Balbinot}, E., {et~al.} 2018, \apj, 862, 114

\bibitem[{{Siegel} {et~al.}(2001){Siegel}, {Majewski}, {Cudworth}, \&
  {Takamiya}}]{siegeletal2001}
{Siegel}, M.~H., {Majewski}, S.~R., {Cudworth}, K.~M., \& {Takamiya}, M. 2001,
  \aj, 121, 935

\bibitem[{{Sollima} {et~al.}(2018){Sollima}, {Mart{\'\i}nez Delgado},
  {Mu{\~n}oz}, {Carballo-Bello}, {Valls-Gabaud}, {Grebel}, {Santana},
  {C{\^o}t{\'e}}, \& {Djorgovski}}]{sollimaetal2018}
{Sollima}, A., {Mart{\'\i}nez Delgado}, D., {Mu{\~n}oz}, R.~R., {et~al.} 2018,
  \mnras, 476, 4814

\bibitem[{{Sollima} {et~al.}(2011){Sollima}, {Mart{\'{\i}}nez-Delgado},
  {Valls-Gabaud}, \& {Pe{\~n}arrubia}}]{sollimaetal2011}
{Sollima}, A., {Mart{\'{\i}}nez-Delgado}, D., {Valls-Gabaud}, D., \&
  {Pe{\~n}arrubia}, J. 2011, \apj, 726, 47

\bibitem[{{Vanderbeke} {et~al.}(2015){Vanderbeke}, {De Propris}, {De Rijcke},
  {Baes}, {West}, \& {Blakeslee}}]{vanderbekeetal2015}
{Vanderbeke}, J., {De Propris}, R., {De Rijcke}, S., {et~al.} 2015, \mnras,
  450, 2692

\bibitem[{{Vanderplas} {et~al.}(2012){Vanderplas}, {Connolly}, {Ivezi{\'c}}, \&
  {Gray}}]{astroml}
{Vanderplas}, J., {Connolly}, A., {Ivezi{\'c}}, {\v Z}., \& {Gray}, A. 2012, in
  Conference on Intelligent Data Understanding (CIDU), 47 --54

\bibitem[{{Webb} {et~al.}(2013){Webb}, {Harris}, {Sills}, \&
  {Hurley}}]{webbetal2013}
{Webb}, J.~J., {Harris}, W.~E., {Sills}, A., \& {Hurley}, J.~R. 2013, \apj,
  764, 124

\bibitem[{{Webb} {et~al.}(2014){Webb}, {Sills}, {Harris}, \&
  {Hurley}}]{webbetal2014}
{Webb}, J.~J., {Sills}, A., {Harris}, W.~E., \& {Hurley}, J.~R. 2014, \mnras,
  445, 1048

\bibitem[{{Yepez} {et~al.}(2019){Yepez}, {Arellano Ferro}, {Schr{\"o}der},
  {Muneer}, {Giridhar}, \& {Allen}}]{yepezetal2019}
{Yepez}, M.~A., {Arellano Ferro}, A., {Schr{\"o}der}, K.~P., {et~al.} 2019,
  \na, 71, 1

\end{thebibliography}


\end{document}